\documentclass[twocolumn]{aastex62}

\newcommand{\dd}{{\rm d}}
\usepackage{ulem}
\usepackage{amsmath}
\usepackage{xcolor} 
\newcommand{\add}[1]{{#1}}

\newcommand{\da}[2]{{{#2}}} 
\usepackage{tabularx}
\usepackage{booktabs}

\graphicspath{{./}{figures/}}

\submitjournal{ApJ}

\shorttitle{A new universal relation}
\shortauthors{Y. Gao et al.}

\begin{document}

\title{A tight universal relation between the shape eccentricity and the moment of inertia for rotating neutron stars}

\correspondingauthor{Lijing Shao}
\email{lshao@pku.edu.cn}

\author[0000-0003-1390-5477]{Yong Gao}
\affiliation{Department of Astronomy, School of Physics, Peking University, Beijing 100871, China}
\affiliation{Kavli Institute for Astronomy and Astrophysics, Peking University, Beijing 100871, China}

\author[0000-0002-1334-8853]{Lijing Shao}
\affiliation{Kavli Institute for Astronomy and Astrophysics, Peking University, Beijing 100871, China}
\affiliation{National Astronomical Observatories, Chinese Academy of Sciences, Beijing 100012, China}

\author[0000-0002-1614-0214]{Jan Steinhoff}
\affiliation{Max Planck Institute for Gravitational Physics (Albert Einstein Institute),
Am M\"uhlenberg 1, Potsdam 14476, Germany}



\begin{abstract}

    Universal relations that are insensitive to the equation of state (EoS) are useful in reducing the parameter space when measuring global quantities of neutron stars (NSs). In this paper, we reveal a new universal relation that connects the eccentricity to the radius and moment of inertia of rotating NSs. We demonstrate that the universality of this relation holds for both conventional NSs and bare quark stars (QSs) in the slow rotation approximation, albeit with different relations. The maximum relative deviation is approximately $1\%$ for conventional NSs and $0.1\%$ for QSs. Additionally, we show that the universality still exists for fast-rotating NSs if we use the dimensionless spin to characterize their rotation. The new universal relation will be a valuable tool to reduce the number of parameters used to describe the shape and multipoles of rotating NSs, and it may also be used to infer the eccentricity or moment of inertia of NSs in future X-ray observations.

\end{abstract}

\keywords{dense matter --- methods: numerical --- stars: rotation}

\section{Introduction} 
\label{sec:intro}

Neutron stars (NSs) are the densest stars in the universe, offering a unique laboratory to study supranuclear matter and gravity in the strong-field regime. Currently, the equation of state (EoS) for the cores of NSs is still poorly understood. Many EoS models with varying compositions and states of dense matter have been developed, leading to significantly different predictions of global properties for NSs~\citep{Lattimer:2000nx}. Therefore, observed global properties of NSs, such as the mass and the radius, can be used to constrain EoS models.

Despite the fact that the global properties of NSs depend sensitively on the EoS models, there exist EoS-insensitive relations that connect various quantities of NSs. These relations are said to be universal because they are insensitive to EoS models to a high degree of accuracy. 
For instance, a universal relation connecting the frequency and damping time of the quadrupolar $f$ mode to the mass and moment of inertia of NSs was discovered by \citet{Lau:2009bu}. \citet{Yagi:2013awa} found the famous I-Love-Q relation for slowly rotating NSs, which links the mass, the moment of inertia, the tidal Love number, and the spin-induced quadrupole moment. Universal relations for NSs are of great significance in both astrophysics and fundamental physics. By providing EoS-insensitive connections between different quantities, these relations allow us to extract global properties of NSs with higher accuracy, and help us study the inverse problem of determining the EoS. The universal relations are also a probe for non-perfect fluid inside NSs. For instance, it has been shown that anisotropic pressure~\citep{Yagi:2015hda}, strong magnetic fields~\citep{Haskell:2013vha}, and ultra-high elasticity~\citep{Lau:2017qtz,Lau:2018mae} can affect the global structure of NSs and potentially break the I-Love-Q universal relation. Moreover, universal relations can break the degeneracy between gravity theories and the uncertainties in EoS, making NSs the ideal laboratories to test gravity \citep{Shao:2022koz}. We refer readers to \citet{Yagi:2013bca}, \citet{Doneva:2017jop}, and references therein for reviews.
 
The exploration of the universal relations for rotating NSs has garnered lots of attention since the discovery of the I-Love-Q relation. The calculations of the I-Q relation were quickly extended to fast rotation by \citet{Doneva:2013rha}, it was shown that the universality of the relation is lost and becomes increasingly EoS-dependent as the spin frequency increases. However, \citet{Pappas:2013naa} and \citet{Chakrabarti:2013tca} demonstrated that the I-Q relation remains universal if dimensionless quantities are used to characterize the spin amplitude instead of the spin frequency $f$. \citet{Pappas:2013naa} and \citet{Yagi:2014bxa} discovered that the first four multipole moments of rotating NSs are universal to some extent. This relation allows for a more accurate description of the spacetime geometry around a NS with fewer parameters. Additionally, \citet{Luk:2018xmt} found another universal relation connecting the radius and orbital frequency of the innermost stable circular orbit (ISCO) to the mass and spin frequency of rotating NSs.

\da{Apart from multipoles and ISCO, the eccentricity is another important parameter to describe rotating NSs. The oblateness induced by 
rotation has a large impact on the X-ray emissions from the surface of X-ray pulsars. To study this effect, \mbox{\citet{Morsink:2007tv}}
parameterized the oblate shape with the compactness of NSs and discovered that the geometric effect induced by the oblateness can 
rival the Doppler effect in certain configurations. To reduce the parameter space of X-ray modeling, \mbox{\citet{Baubock:2013gna}} derived a 
universal relation between the eccentricity and the compactness of slowly rotating NSs. \mbox{\citet{AlGendy:2014eua}} also found 
an EoS-insensitive fit of the eccentricity, using a slightly different parametrization for the surface other than the one used 
in the Hartle-Thorne formalism.}{Apart from multipoles and the ISCO, the oblate shape is also important. In the canonical pulse-profile modelling of X-ray pulsars, photons are emitted from an oblate surface and assumed to propagate in a Schwarzschild background, which is the so-called ``Oblate $+$ Schwarzschild'' (O $+$ S) approximation. To give an analytical formula to describe the rotation-induced oblateness, \citet{Morsink:2007tv} and \citet{AlGendy:2014eua} parametrized the oblate shape withe the spin frequency and compactness of NSs. Recently, \citet{Silva:2020oww} developed a more accurate fitting formula compared to \citet{Morsink:2007tv,AlGendy:2014eua}, which better describes the large deformation of the surface for very rapid rotation. These fitting formulas can capture the shape of NSs at a wide range of spin frequencies, compactnesses, and EoSs (i.e., universal to some extent). In slow rotation, \citet{Baubock:2013gna} also explored the universal relation of rotating NSs. They showed that both the moment of inertia and the surface eccentricity can be approximately represented by a single parameter, the compactness.} \citet{Frieben:2012dz}uncovered quasi-universal relations relating surface distortion to spin frequency and magnetic field, which can be used to calculate surface distortion up to significant levels of rotation and magnetization.

In this paper, we discover a new universal relation between the surface eccentricity and the moment of inertia for rotating NSs. The paper is structured as follows. In Sec.~\ref{sec:slow}, we provide a definition of multipoles and eccentricity in the slow rotation approximation and present the universal relation for both conventional NSs and QSs. In Sec.~\ref{sec:fast}, we investigate the universal relation for fast rotating NSs. Discussion of possible applications and connections of the new universal relation to early work is shown in Sec.~\ref{sec:discussion}. Throughout the paper, we use geometric units with $G=c=1$.

\section{A new universal relation in the slow-rotation approximation}
\label{sec:slow}

\subsection{Multipole moments and shape parameters}

To study the universal relation, we first give an overview of the
structures and shape parameters of slowly rotating NSs.
Following \citet{Hartle:1967he} and \citet{Hartle:1968si}, we construct these 
stars by solving the Einstein equations perturbatively in a slow-rotation expansion to quadratic order in the 
spin. At the zeroth order in spin, we obtain the mass $M$ and the radius $R$ of the non-rotating background.
At the first order in spin, we extract the angular momentum $J$, from which we can define the moment of inertia $I$ and the 
dimensionless spin $\chi$ as 
    $I\equiv {J}/{\Omega}$ and $\chi \equiv {J}/{M^2}$,
where $\Omega$ is the angular frequency of the rotating star. 
Universal relations usually connect dimensionless quantities. The dimensionless moment of inertia $\bar I$ is usually  
defined as 
    ${\bar I} \equiv {I}/{M^3}$.
At the second order in spin, the star is deformed into an oblate shape, and we get the spin-induced quadrupole moment $Q\equiv -{J^{2}}/{M}-{8}K M^{3}/5$.
The parameter $K$ depends on the EoS of NSs and equals to zero for Kerr black holes according to the no hair theorem. 
The dimensionless quadrupole moment is defined as 
    $\bar Q\equiv -{Q}/{M^{3}\chi^{2}}$.
The I-Q relation connects the dimensionless quantities $\bar I$ and $\bar Q$.
The exterior spacetime of a slowly rotating NS can be fully described up to the quadratic order in spin by 
the mass $M$, the angular momentum $J$, and the quadrupole moment $Q$~\citep{Hartle:1968si}. 

Observationally, some observation of a rotating NS depends on the geometry of its surface.
We use the eccentricity $e_{\rm s}$ to describe the oblate shape of a NS, 
\begin{equation}
    \label{eqn:def_eccentricity}
    {e}_{\rm s}\equiv \sqrt{\left(\frac{R_{\mathrm{\rm eq}}}{R_{\mathrm{\rm p}}}\right)^{2}-1}\,,
\end{equation}
where $R_{\rm eq}$ and $R_{\rm p}$ are the equatorial and polar radii in a specific coordinate.
In the Hartle-Thorne coordinate, the isodensity surface at radial coordinate $r$ in the non-rotating star is displaced to
\begin{equation}
    \label{eqn:ht_surface}
    r \to r+\xi_{0}(r)+\xi_{2}(r) P_{2}(\cos \theta) \,,
\end{equation}
in the rotating configuration, where $\xi_{0}$ and $\xi_{2}$ are spherical and quadrupole displacements respectively, and 
$P_{2}(\cos \theta)$ is the Legendre polynomial. 
Combining Eqs.~(\ref{eqn:def_eccentricity}--\ref{eqn:ht_surface}), we get the surface eccentricity in the Hartle-Thorne coordinate as 
\begin{equation}
    \label{eqn:htshape}
    e_{\rm HT}=\left[-3\left(\xi_{2}(R) / R\right)\right]^{1 / 2}\,.
\end{equation}

Equation~(\ref{eqn:ht_surface}) describes the isodensity surface in a particular coordinate system. By embedding the isodensity surface into a three-dimensional flat space (denoted by $r^{*}$, $\theta^{*}$, $\phi^{*}$), \citet{Hartle:1968si} found an invariant parametrization of the oblate surface. To the second order of the spin, the desired surface is a spheroid with 
\begin{align}
    r^{*}\left(\theta^{*}\right)&=r+\xi_{0}(r)\nonumber \\
     &+\Big\{\xi_{2}(r)+r\left[v_{2}(r)-h_{2}(r)\right]\Big\} P_{2}\left(\cos \theta^{*}\right)\,.
\end{align}
Here $v_2$ and $h_2$ are metric functions at the second order in spin. The eccentricity of the stellar surface embedded in flat space is then given by
\begin{equation}
    \label{eqn:flatshape}
    e_{*}=\big\{-3\big[v_{2}(R)-h_{2}(R)+\xi_{2}(R)/ R\big]\big\}^{1 / 2}\,,
\end{equation}
where the superscript ``$*$'' denotes the eccentricity observed in the flat space.

\subsection{A universal relation for the eccentricity of NSs}

The universal relation that we discovered connects the quantity $e_{\rm s}/R\Omega$ and the dimensionless 
moment of inertia $\bar I$. 
Samiliar to the I-Q relation~\citep{Yagi:2013awa} and the three hair relation for the multipole moments~\citep{Yagi:2014bxa},
the normalization factors $M$ and $R$ are quantities of the non-rotating background in the slow-rotation approximation. For convenience, we define 
a dimensionless radius, $\hat{R}\equiv{R\Omega}$. We have verified that the universal relation exists for both the eccentricity in the Hartle-Thorne coordinate, $e_{\rm HT}$, and the eccentricity of the embedding surface, $e_{*}$. In the following, we use $e_{\rm *}$ to illustrate the results.

\begin{figure}
    \centering 
    \includegraphics[width=8.5cm]{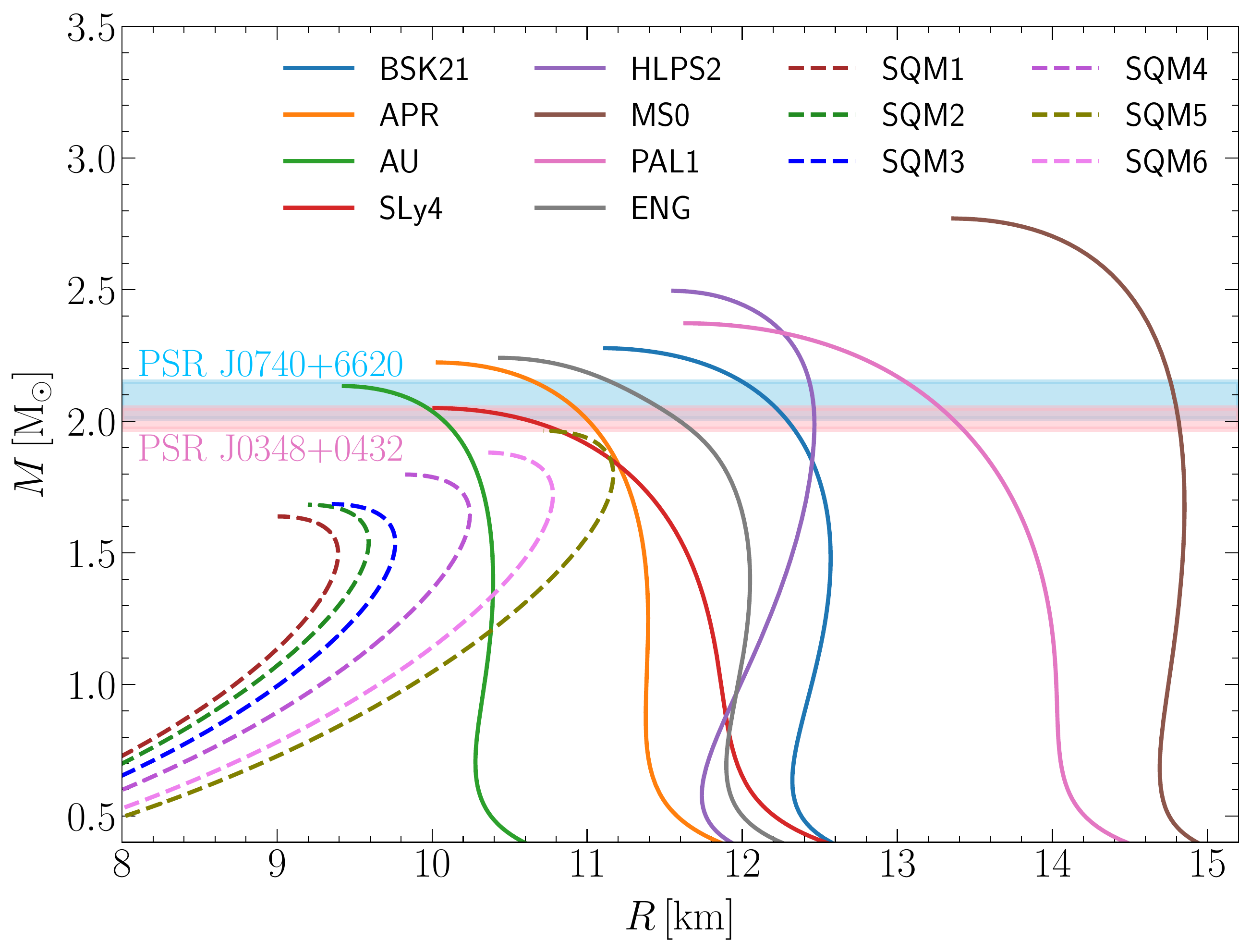}
    \caption{The mass-radius relation for selected EoS models of NSs ({solid}) and QSs ({dashed}). The $1$-$\sigma$ regions of the
    mass measurements of PSR~J0348$+$0432 \citep{Antoniadis1233232} and PSR~J0740$+$6620 \citep{Fonseca:2021wxt} are illustrated.}
    \label{fig:mr}
\end{figure}

We first study the universal relation for conventional NSs. Our selection of realistic EoSs includes BSK21~\citep{Goriely:2010bm}, AU~\citep{Wiringa:1988tp}, HLPS~\citep{Hebeler:2013nza}, PAL1~\citep{Prakash:1988md}, APR~\citep{Akmal:1998cf}, SLy4~\citep{Douchin:2001sv}, MS0~\citep{Mueller:1996pm}, and ENG~\citep{Engvik1994}. As shown in Fig.~\ref{fig:mr}, these models cover a wide range in the mass-radius diagram of static NSs, and all of them have maximal NS mass larger than $2\,M_{\odot}$. Although the very stiff EoSs MS0 and PAL1 have been ruled out by the tidal deformability from GW170817~\citep{GW170817}, we include them to demonstrate that the universality exists for a large family of EoSs.  

\begin{figure}
    \centering 
    \includegraphics[width=8.5cm]{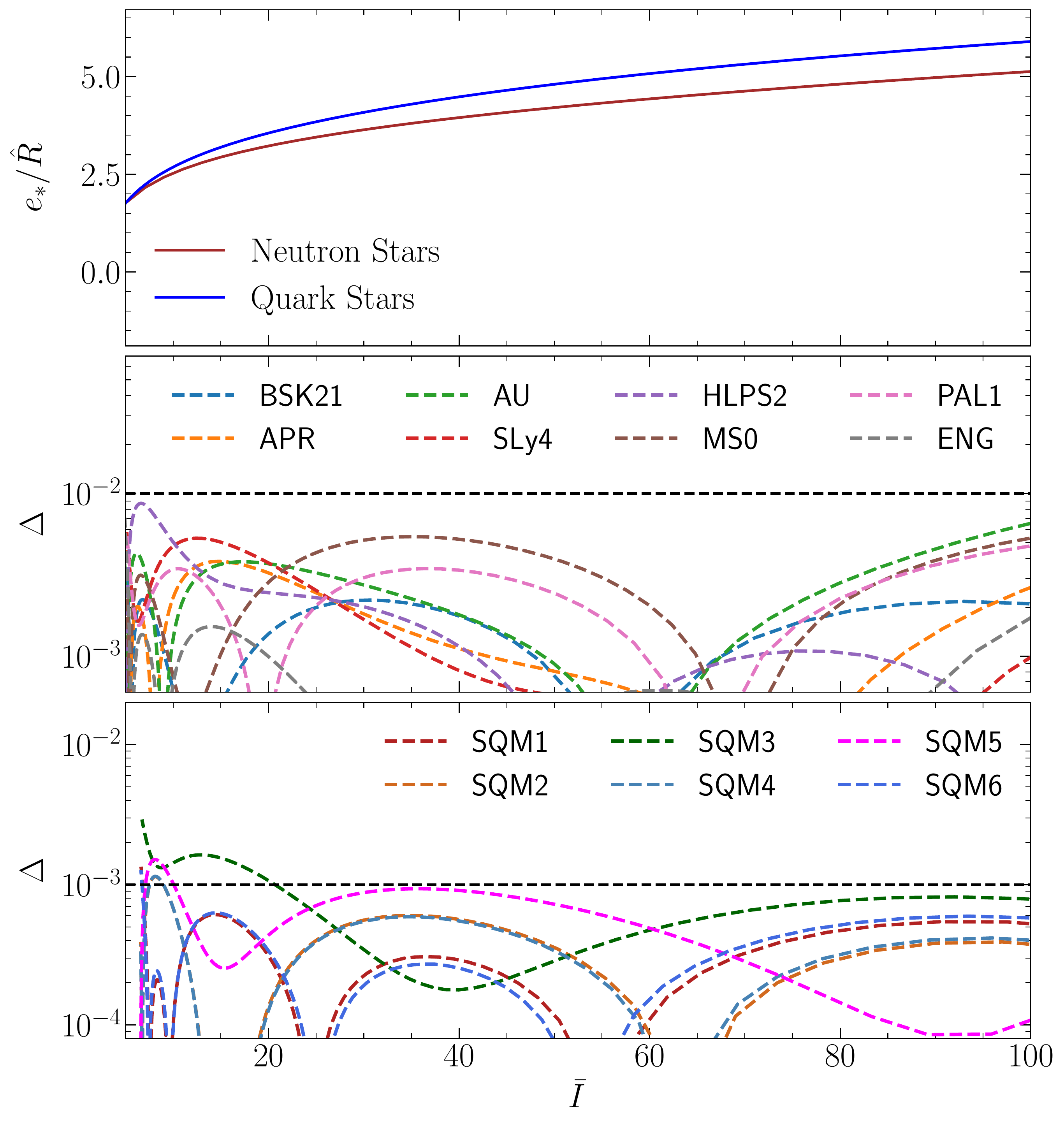}
    \caption{The $e_{\rm *}/\hat{R}$-$\bar I$ universal relation for slowly-rotating models. The upper panel shows the fitted universal relations for both NSs and QSs. The middle (lower) panel presents the relative deviation for NSs (QSs).}
    \label{fig:universal_e}
\end{figure}

Our universal relation is described with great accuracy by 
\begin{equation}
    \label{eqn:fitting}
    \frac{e_{*}}{\hat R}=\sum_{k=0}^3 a_{ k}\left(\ln \bar I\right)^k\,,
\end{equation}
where $a_{ k}$'s are fitting coefficients with $a_0 =-0.855572$, $a_1 = 2.185502$, $a_2 =-0.428061$, $a_3 = 0.051177$. 
Note that we use the dimensionless moment of inertia $\bar I \leq 100$, which corresponds to $M \gtrsim 0.5\,M_{\odot}$ for selected models.
We define the relative deviation to Eq.~(\ref{eqn:fitting}) as 
\begin{equation}
    \Delta=\frac{e_{*}/ \hat{R}-(e_{*}/ \hat{R})_{\rm fit}}{(e_{*}/ \hat{R})_{\rm fit}} \,.
\end{equation}
As shown in Fig.~\ref{fig:universal_e}, the relative deviation to the universal relation is smaller than $1\%$ for selected models of EoSs.   NSs with $M\gtrsim 1\,M_{\odot}$ are more relevant for astrophysics, and in this case, the dimensionless moment of inertia $\bar{I}\lesssim 30$ for selected EoS models, and the universal relation takes a simpler form,
\begin{equation}
    \frac{e_{*}}{\hat R} = 0.11418+1.04115\ln \bar I\,.
\end{equation}
The relative deviation to this relation is less than $\sim 1\%$.

For QSs, we use the phenomenological MIT bag model to describe the quark matter. This model assumes a nearly equal number of $u$, $d$, $s$ quarks and a small fraction of electrons confined within a bag of vacuum energy density $B$ \citep{Farhi:1984qu,Witten:1984rs}. We account for the mass  of the $s$ quark, $m_{\rm s}$, and include the quark-gluon interaction to the lowest order in $\alpha_c=g^2/4\pi$. To investigate the universal relation for QSs, we employ six different EoSs with varying combinations of $m_{\rm s}$, $\alpha_{\rm c}$, and $B$ in Table~\ref{tab:sqm}. The resulting mass-radius relation is displayed in Fig.~\ref{fig:mr}. 

\begin{table}
    \caption{Parameters for QSs in the MIT bag model.}
    \centering
    \begin{tabular}{lccc}
    \toprule
        Model & $B\,(\rm MeV\,fm^{-3})$  & $m_{\rm s}\,(\rm MeV)$ & $\alpha_{\rm c}$ \\ \hline
        SQM1 & 80 & 100 & 0 \\ 
        SQM2  & 80 & 50 & 0.1 \\ 
        SQM3  & 70 & 150 & 0 \\ 
        SQM4  & 70 & 50 & 0.3 \\ 
        SQM5  & 60 & 0 & 0 \\ 
        SQM6  & 60 & 100 & 0.4 \\ \hline
    \end{tabular}
    \label{tab:sqm}
\end{table}

The relation between $e_{*}/\hat{R}$ and $\bar I$ is different from that of NSs, but it is still universal and can be well fitted by 
\begin{equation}
    \label{eqn:qs_fitting}
    \frac{e_{*}}{\hat R}=\sum_{k=0}^4 b_{ k}\left(\ln \bar I\right)^k\,,
\end{equation}
with coefficients $b_0= -1.499749$, $b_1= 2.911859$, $b_2 =-0.749237$, $b_3 = 0.137057$, and $b_4 =-0.007801$. Interestingly, the deviation from the universal relation for QSs is much smaller than that for NSs, with relative deviation less than $\sim 0.1\%$, as shown in the lower panel of Fig.~\ref{fig:universal_e}. For QSs in our study, the condition for $M \gtrsim 1\,M_{\odot}$ corresponds to $\bar I \lesssim 20$. Within this range, the universal relation can be approximated by a simpler fitting formula, 
\begin{equation}
    \frac{e_{*}}{\hat R} = -0.043372 + 1.210546\ln(\bar{I} -0.470579)\,,
\end{equation}
The relative deviation from this fitting formula is less than $\sim 0.3\%$.

\add{Compared to the parametrization in \citet{Baubock:2013gna}, the new universal relation that we propose incorporates an extra parameter, namely the moment of inertia, in addition to the parameters $R$, $M$, and $\Omega$. But the new universal relation is much tighter than that of \citet{Baubock:2013gna}.}

\section{Universal relation for fast rotating NSs}
\label{sec:fast}

Fast rotation is relevant for sub-millisecond pulsars, nascent NSs after supernovae, and NSs formed in binary NS mergers. Rapid rotation causes NSs to develop a more obvious oblate shape. In this section, we explore the universal relation for rapidly rotating NSs using the {\tt RNS} code developed by \citet{Stergioulas:1994ea}. 

The {\tt RNS} code uses a quasi-isotropic coordinate system to represent the line element of the stationary axisymmetric spacetime,
\begin{align}
\dd s^2= & -e^{2 \nu} \dd t^2+ B^2 r^2 \sin ^2 \theta  e^{-2 \nu}(\dd \phi-\omega \dd t)^2 \nonumber \\
& +e^{2(\xi-\nu)}\left(\dd r^2+r^2 \dd \theta^2\right),
\end{align}
where $\nu$, $B$, $\omega$, and $\xi$ are metric functions that depend on $r$ and $\theta$. Assuming a perfect fluid and uniform rotation, we obtain the stellar structure and spacetime metric. The conserved angular momentum $J$ can be computed from a volume integration over the matter field. The moment of inertia $I$ and the dimensionless spin have the same definition as before. The quadrupole moment $Q$ can be obtained from the asymptotic expansion of the metric functions. \da{The surface eccentricity is formally given by Eq.~(\ref{eqn:def_eccentricity}), with the eccentricity, equatorial and polar radii defined in the quasi-isotropic coordinate. We use the notation $e_{i}$ to denote the eccentricity in the quasi-isotropic coordinate.}{The surface eccentricity is formally given by Eq.~(\ref{eqn:def_eccentricity}), with the eccentricity $e_{i}$, equatorial radius $R_{\rm eq}^{i}$, and polar radius $R_{\rm q}^{i}$ defined in the quasi-isotropic coordinate.} Note that, unlike in the slow-rotation approximation, the normalization factor $M$ is the mass for the rotating configuration, and $\hat{R}\equiv R_{\rm eq}^{i}\Omega$.

\begin{figure}
    \centering 
    \includegraphics[width=8.5cm]{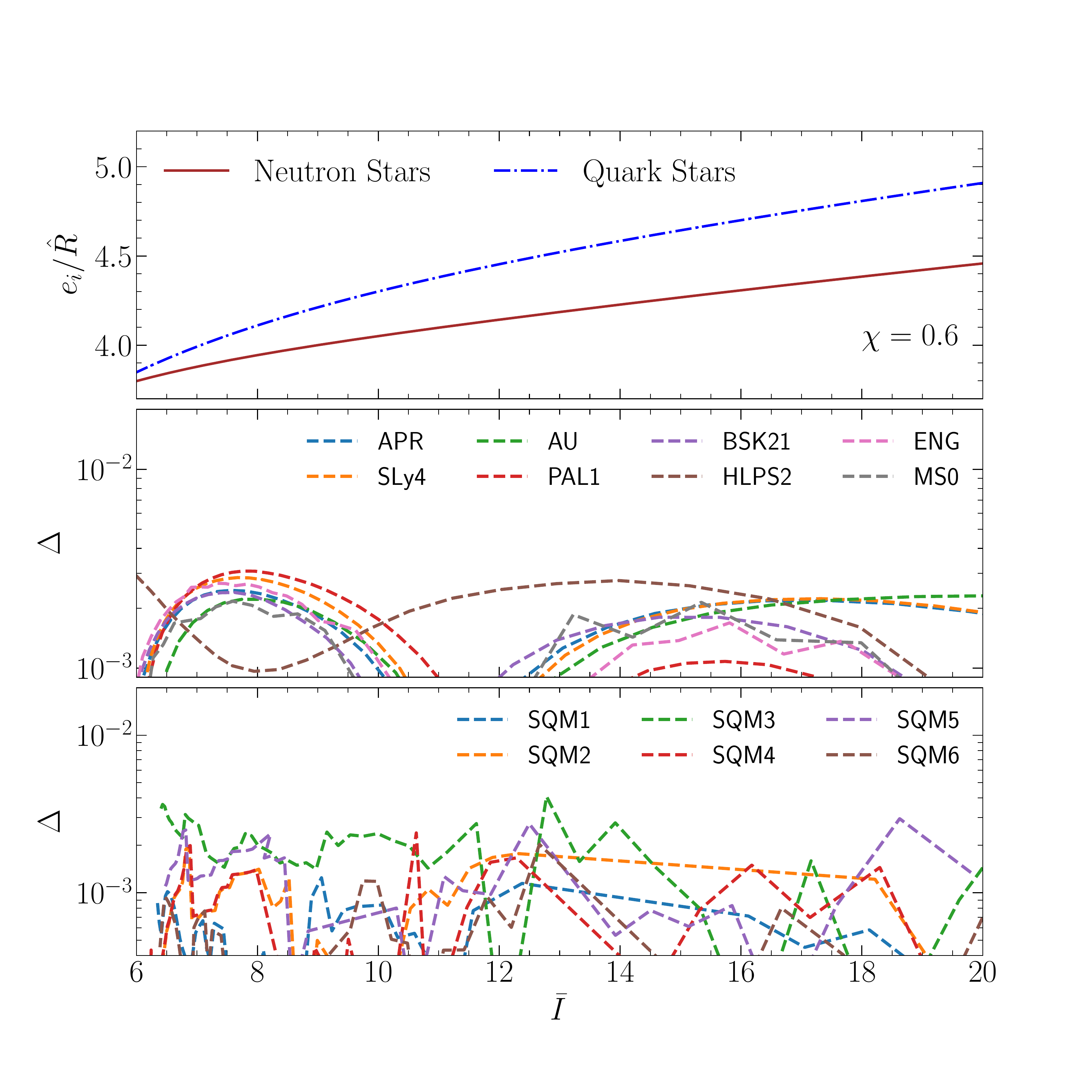}
    \caption{\add{The $e_{i}/\hat{R}$-$\bar I$ universal relation for fast-rotating models with the dimensionless spin $\chi=0.6$. The upper panel shows the fitted universal relations for both NSs and QSs. The middle (lower) panel presents the relative deviation for NSs (QSs).}}
    \label{fig:universal_fast1}
\end{figure}

To study universal relations for rapidly rotating NSs, it is necessary to use a suitable parameter to characterize their spin amplitude. As demonstrated by \citet{Doneva:2013rha}, if one uses the spin frequency $f$ as the parameter, the I-Q relation for fast rotating NSs is lost. Similarly, the universal relation that we discovered also breaks down for fixed spin frequencies. However, \citet{Pappas:2013naa} and \citet{Chakrabarti:2013tca} found that the I-Q relation is still universal for fast rotating NSs if one chooses dimensionless spin parameters such as $\chi$, $Mf$, and $Rf$, instead of the dimensionful $f$. Inspired by their work, we use $\chi$ to characterize the spin amplitude and find that the $e_{i}/\hat{R}$-$\bar I$ relation for both conventional NSs and strange QSs is still universal.

\begin{figure}
    \centering 
    \includegraphics[width=8cm]{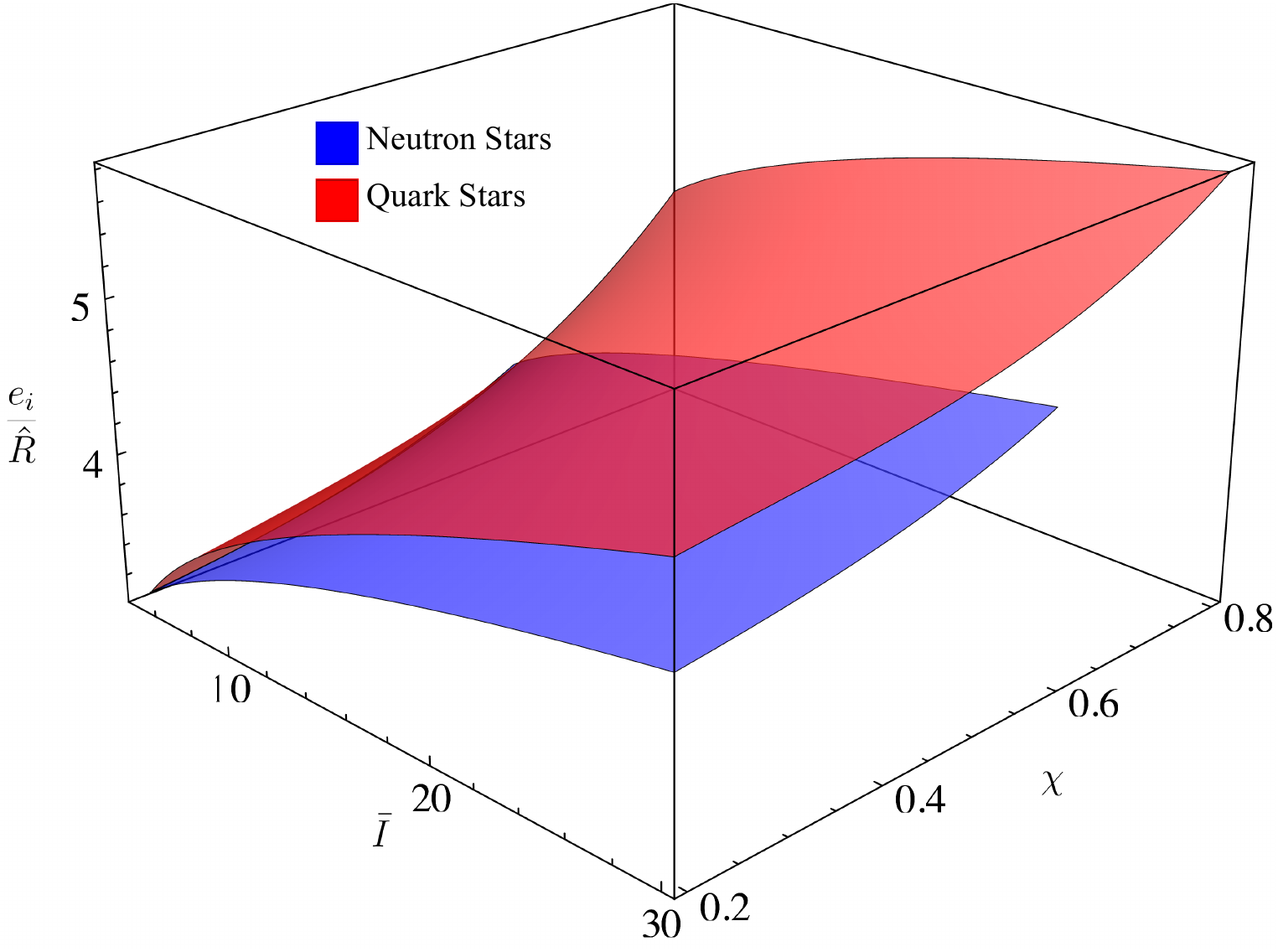}
    \caption{\add{The surface of the fitting formula Eq.~(\ref{eqn:coefficient}) for NSs (blue) and QSs (red). Models of NSs and QSs fall on two well-defined surfaces.}}
    \label{fig:universal_fast2}
\end{figure}

According to \citet{Lo:2010bj}, the maximum value of the dimensionless spin parameter $\chi$ for NSs rotating at the Keplerian frequency is about 0.7 for various EoS models. This limit is nearly independent of the mass of the NS if the mass is larger than $1\,M_{\odot}$. However, for QSs in the MIT bag model, the spin parameter can be larger than unity and does not have a universal upper limit. Its value also depends strongly on the bag constant and the mass of the star. \da{Therefore, in Fig.~\ref{fig:universal_fast1}, we display three representative cases with $\chi=0.4$ and $0.6$ for conventional NSs. The relative deviation from the universal relation is less than $\sim 1\%$. The cases for QSs with $\chi=0.6$ and $\chi=0.8$ are shown in Fig.~\ref{fig:universal_fast2}, and the relative deviation is on the order of $0.1\%$, which is again much tighter than conventional NSs.}{Therefore, we generate data points in the regime $0.2\leq \chi \leq 0.6$ for NSs and $0.2\leq \chi \leq 0.8$ for QSs.}

\add{In Fig.~\ref{fig:universal_fast1}, we show the universal relation for both NSs and QSs with $\chi=0.6$. The relative errors are on the order of $0.3\%$. More explicitly, the $e_{\rm i}/\hat{R}$-$\bar{I}$ relation with a dependence on $\chi$ can be fitted by}
\begin{equation}
    \label{eqn:coefficient}
    \add{\frac{e_{\rm i}}{\hat{R}}=\sum_{i, j} \mathcal{A}_{i j} \chi^i \log ^j \bar{I}\,,}
\end{equation}
\add{where the numerical coefficients $\mathcal{A}_{i j}$ for NSs and QSs are given in Table \ref{tab:coefficients}. The maximum relative errors of the fitting formula are on the order of $1\%$ for NSs and $0.3\%$ for QSs. In Fig.~\ref{fig:universal_fast2}, we present the surfaces described by the fitting formula Eq.~(\ref{eqn:coefficient}). For a given $\chi$ and $\bar I$, the value of $e_{\rm i}/\hat{R}$ for QSs is always larger than that for NSs. At the maximum mass limit, the two surfaces become closest.}
\def\arraystretch{1.1}
\begin{table}
	\caption{\add{Numerical coefficients for the two-parameter fitting formula Eq.~(\ref{eqn:coefficient}).}}
	\centering 
	\begin{tabular}{lcccc}
		\toprule
		$i=$&
		$0$ & 
		$1$ & 
		$2$ & 
		$3$ \\ 
		\midrule
        \multicolumn{5}{c}{Coefficients for neutron stars} \\
        \cmidrule(lr){1-5}
        $\mathcal{A}_{i 0}$ & $2.149039$ & $-0.146205$ & $-1.867329$ & $5.65915$ \\ 
        $\mathcal{A}_{i 1}$ & $1.23962$ & $2.21818$ & $0.942935$ & $-2.87271$ \\
        $\mathcal{A}_{i 2}$ & $-0.165024$ & $-2.63504$ & $2.28603$ & $-2.53576$ \\ 
        $\mathcal{A}_{i 3}$ & $0.158013$ & $1.04597$ & $-1.80829$ & $2.23285$ \\

        \midrule
        \multicolumn{5}{c}{Coefficients for quark stars} \\
        \cmidrule(lr){1-5}
        $\mathcal{A}_{i 0}$ & 0.98273 & $-5.06930$ & 12.9114 & 5.65915 \\ 
        $\mathcal{A}_{i 1}$ & 4.29319 & 16.3498 & $-48.8296$ & 30.4711 \\
        $\mathcal{A}_{i 2}$ &  $-3.12737$ & $-12.9528$ & 50.1446 & $-37.8440$ \\ 
        $\mathcal{A}_{i 3}$ & 1.30675 & 3.27539 & $-16.5473$ & 13.8336 \\
		\bottomrule
	\end{tabular}
	\label{tab:coefficients}
\end{table}

\add{In our study, we have defined the eccentricity in the quasi-isotropic coordinate system, where the radial coordinate $r$ corresponds to the isotropic Schwarzschild coordinate in the limit of zero spin. However, \cite{Morsink:2007tv}, \cite{AlGendy:2014eua}, and \cite{Silva:2020oww} define the eccentricity differently. We know that circles centered about the symmetric axis have circumference $2\pi \bar{r}$, where $\bar{r}$ is related to $r$ and $\theta$ by }
\begin{equation}
   \add{\bar{r}=e^{-\nu(r,\theta)}B(r,\theta)r\sin\theta=r_{c}(r,\theta)\sin\theta\,.}
\end{equation}
\add{Here $r_{c}$ corresponds to the Schwarzschild coordinate in the limit of zero spin. By using $r_{c}$, the equatorial and polar radii can be defined as }
\begin{equation}
   \add{ R_{\rm eq}^{c}=r_{c}(R_{\rm eq}^{i},\theta=\frac{\pi}{2})\,,\quad R_{\rm p}^{c}=r_{c}(R_{\rm p}^{i},\theta=0)\,,}
\end{equation}
\add{where $R_{\rm eq}^{c}$ is the circumferential radius of the star in the equatorial plane. Then the surface eccentricity $e_{c}$ can be obtained from Eq.~(\ref{eqn:def_eccentricity}).}

\begin{figure}
    \centering 
    \includegraphics[width=8.5cm]{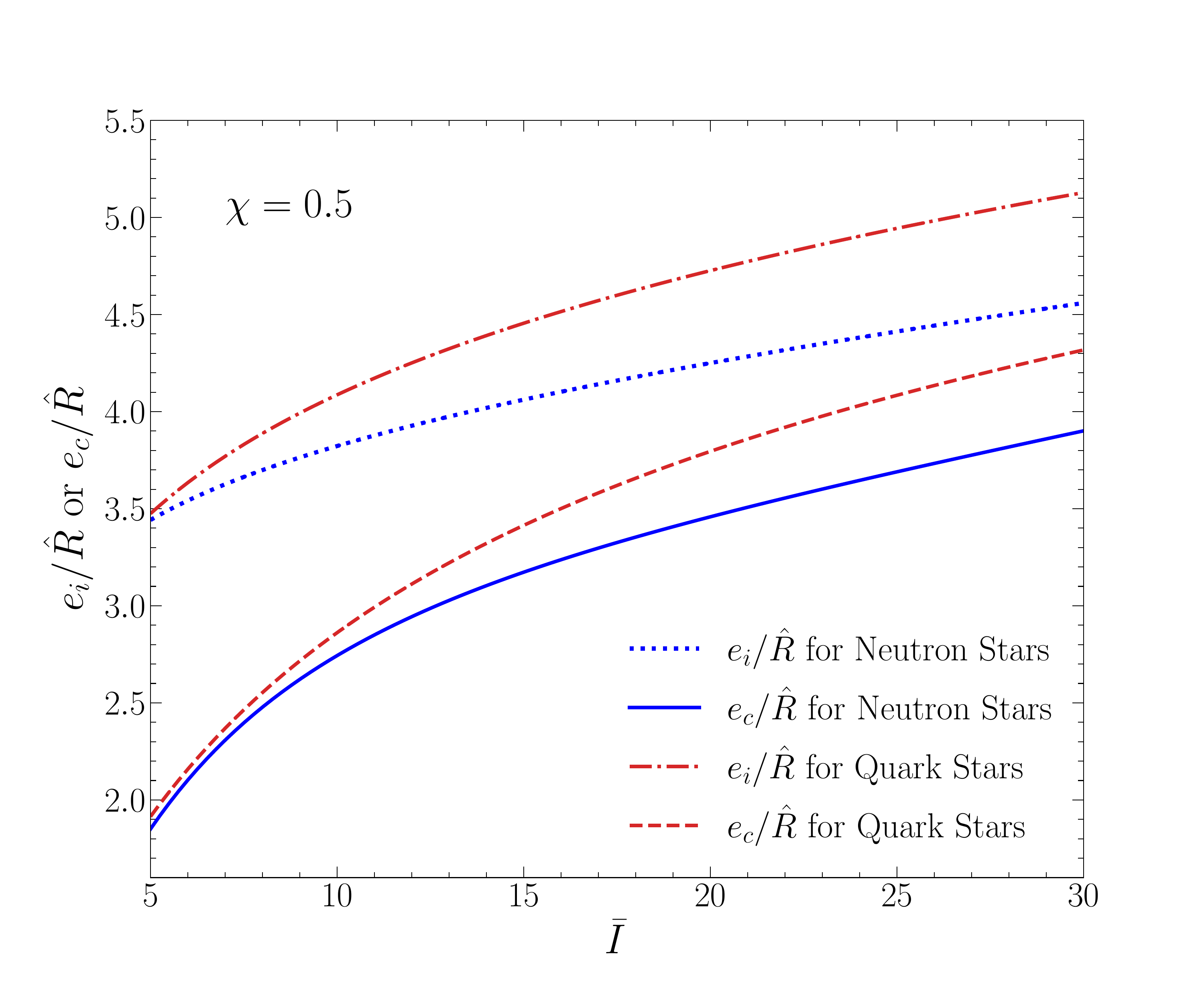}
    \caption{\add{The $e_{i}/\hat{R}$-$\bar{I}$ and $e_{c}/\hat{R}$-$\bar{I}$ relations for NSs ({\it blue}) and QSs ({\it red}). Here we take the dimensionless spin $\chi=0.5$.}}
    \label{fig:compare}
\end{figure}

\add{We have found that the universality still holds for the eccentricity $e_{c}$, but its relation differs from that of $e_{i}$. Specifically, for given EoS families and a fixed value of $\bar I$, the value of $e_{i}/\hat{R}$ is larger than $e_{c}/\hat{R}$. To compare the universal relations for these two eccentricities, we have included an example in Fig.~\ref{fig:compare}. The figure shows the $e_{i}/\hat{R}$-$\bar I$ and $e_{c}/\hat{R}$-$\bar I$ relations for NSs and QSs at $\chi=0.5$, which highlights the differences of the universal relations for these two eccentricities.}

\section{Discussion}
\label{sec:discussion}

\da{The relation between $e_{\rm s}/\hat{R}$ and $\bar I$ adds a new tight universal relation to the known ones. Here, $e_{\rm s}$ is the surface eccentricity formally defined in Eq.~(\ref{eqn:def_eccentricity}) without assigning a specific coordinate system. In our paper, it includes $e_{\rm s}^{\rm HT}$, $e_{\rm s}^*$, and $\tilde e_{\rm s}$, which are eccentricities defined in different coordinate systems. The effect of gauge choice on the eccentricity is very small. All of these eccentricities satisfy the universal relation very well.}{The universal relation between $e_{\rm s}/\hat{R}$ and $\bar I$ adds a new tight universal relation to the known ones. Here, $e_{\rm s}$ is the surface eccentricity formally defined in Eq.~(\ref{eqn:def_eccentricity}), including $e_{\rm HT}$, $e_{*}$, $e_{i}$, and $e_{c}$ in our work. It is important to note that these different definitions of eccentricity lead to different universal relations.} Since the I-Q relation~\citep{Yagi:2013awa,Chakrabarti:2013tca} connects $\bar I$ to $\bar Q$, and the three-hair relation~\citep{Pappas:2013naa,Yagi:2014bxa} connects $\bar Q$ to two other higher-order multipoles, the relation between $e_{\rm s}/\hat{R}$ and these normalized multipole moments is also universal. Combined with the I-Love relation~\citep{Yagi:2013awa}, we have a universal relation between $e_{\rm s}/\hat{R}$ and the dimensionless tidal Love number. Moreover, the universal relation between the $f$-mode oscillation and $\bar I$~\citep{Lau:2009bu} helps us connect $e_{\rm s}/\hat{R}$ to the frequency and damping time of the quadrupolar $f$ mode.


Universal relations are a powerful tool to reduce modelling uncertainties and inferring NS parameters. As we discussed before, the eccentricity of rotating NSs is an observable and is an important input for X-ray modelling. The oblateness induced by rotation at frequencies above $300\,\rm Hz$ produces a geometric effect that has imprints in the pulse profile of X-ray pulsars~\citep{Morsink:2007tv}. For some emission configurations, the oblateness effect can rival the Doppler effect. Besides, the effects of oblateness need to be taken into account when measuring the radii of NSs from rotationally broadened atomic lines~\citep{Baubock:2012bj}. 

\da{On one hand, our new universal relation reduces the number of parameters used to describe the shape and multipoles of rotating NSs. On the other hand, the new universal relation can be used to infer the eccentricity of NSs if the mass, radius, spin frequency, and moment of inertia are inferred in future X-ray observations.}{The new universal relation we obtained can be used to help infer NS properties. For example, if future X-ray observations can measure the eccentricity and radius of pulsars, one can use the universal relation to forecast the moment of inertia of NSs with similar masses, and use the I-Love relation to test gravity~\citep{Silva:2020acr}. Conversely, if the moment of inertia is obtained through observations of, say, the Double Pulsar system~\citep{Lattimer:2004nj,Hu:2020ubl,Kramer:2021jcw} or GWs from a binary NS inspiral~\citep{Lau:2009bu}, our universal relation can be employed to improve the inference of the radii of NSs through X-ray observations.
} 

\add{For very rapid rotation, deviations from a simple ellipse become potentially important. Currently, telescopes like NICER observe only slowly rotating NSs, for which approximating the shape as an ellipse is accurate enough. However, if NICER or similar telescopes were to observe highly rapidly rotating pulsars, the full shape function for the surface would be required. Previous studies, such as \citet{Morsink:2007tv}, \citet{AlGendy:2014eua}, and \citet{Silva:2020oww}, have parameterized the oblate shape of NSs using the spin frequency and compactness as parameters. In comparison to these works, our newly discovered universal relation incorporates an additional parameter, the moment of inertia, and only focuses on the eccentricity of the star. In the future, it's necessary to extend our framework to accurately fit the shape of the star at all latitudes, especially when interpreting the observations of rapidly rotating NSs.
}

\begin{acknowledgments}
We thank \add{the anonymous referee for helpful comments, and Zexin Hu and} Enping Zhou for useful discussions. This work was supported by the National SKA Program of China (2020SKA0120300), the National Natural Science Foundation of
China (11975027, 11991053), the
Max Planck Partner Group Program funded by the Max Planck Society, and the
High-Performance Computing Platform of Peking University.
\end{acknowledgments}

\software{RNS \citep{Stergioulas:1994ea}}

\bibliography{refs}{}

\begin{thebibliography}{}
\expandafter\ifx\csname natexlab\endcsname\relax\def\natexlab#1{#1}\fi
\providecommand{\url}[1]{\href{#1}{#1}}

\bibitem[{Abbott {et~al.}(2017)}]{GW170817}
Abbott, B.~P., {et~al.} 2017, Phys. Rev. Lett., 119, 161101

\bibitem[{Akmal {et~al.}(1998)Akmal, Pandharipande, \&
  Ravenhall}]{Akmal:1998cf}
Akmal, A., Pandharipande, V.~R., \& Ravenhall, D.~G. 1998, Phys. Rev. C, 58,
  1804

\bibitem[{AlGendy \& Morsink(2014)}]{AlGendy:2014eua}
AlGendy, M., \& Morsink, S.~M. 2014, ApJ, 791, 78

\bibitem[{Antoniadis {et~al.}(2013)Antoniadis, Freire, Wex, Tauris, Lynch, van
  Kerkwijk, Kramer, Bassa, Dhillon, Driebe, Hessels, Kaspi, Kondratiev, Langer,
  Marsh, McLaughlin, Pennucci, Ransom, Stairs, van Leeuwen, Verbiest, \&
  Whelan}]{Antoniadis1233232}
Antoniadis, J., Freire, P. C.~C., Wex, N., {et~al.} 2013, Science, 340, 448

\bibitem[{Baub\"ock {et~al.}(2013)Baub\"ock, Berti, Psaltis, \&
  \"Ozel}]{Baubock:2013gna}
Baub\"ock, M., Berti, E., Psaltis, D., \& \"Ozel, F. 2013, ApJ, 777, 68

\bibitem[{Baubock {et~al.}(2013)Baubock, Psaltis, \& Ozel}]{Baubock:2012bj}
Baubock, M., Psaltis, D., \& Ozel, F. 2013, ApJ, 766, 87

\bibitem[{Chakrabarti {et~al.}(2014)Chakrabarti, Delsate, G\"urlebeck, \&
  Steinhoff}]{Chakrabarti:2013tca}
Chakrabarti, S., Delsate, T., G\"urlebeck, N., \& Steinhoff, J. 2014, Phys.
  Rev. Lett., 112, 201102

\bibitem[{Doneva \& Pappas(2018)}]{Doneva:2017jop}
Doneva, D.~D., \& Pappas, G. 2018, Astrophys. Space Sci. Libr., 457, 737

\bibitem[{Doneva {et~al.}(2013)Doneva, Yazadjiev, Stergioulas, \&
  Kokkotas}]{Doneva:2013rha}
Doneva, D.~D., Yazadjiev, S.~S., Stergioulas, N., \& Kokkotas, K.~D. 2013,
  ApJL, 781, L6

\bibitem[{Douchin \& Haensel(2001)}]{Douchin:2001sv}
Douchin, F., \& Haensel, P. 2001, A\&A, 380, 151

\bibitem[{Engvik {et~al.}(1994)Engvik, Hjorth-Jensen, Osnes, Bao, \&
  \O{}stgaard}]{Engvik1994}
Engvik, L., Hjorth-Jensen, M., Osnes, E., Bao, G., \& \O{}stgaard, E. 1994,
  Phys. Rev. Lett., 73, 2650

\bibitem[{Farhi \& Jaffe(1984)}]{Farhi:1984qu}
Farhi, E., \& Jaffe, R.~L. 1984, Phys. Rev. D, 30, 2379

\bibitem[{Fonseca {et~al.}(2021)}]{Fonseca:2021wxt}
Fonseca, E., {et~al.} 2021, ApJL, 915, L12

\bibitem[{Frieben \& Rezzolla(2012)}]{Frieben:2012dz}
Frieben, J., \& Rezzolla, L. 2012, Mon. Not. Roy. Astron. Soc., 427, 3406

\bibitem[{Goriely {et~al.}(2010)Goriely, Chamel, \& Pearson}]{Goriely:2010bm}
Goriely, S., Chamel, N., \& Pearson, J.~M. 2010, Phys. Rev. C, 82, 035804

\bibitem[{Hartle(1967)}]{Hartle:1967he}
Hartle, J.~B. 1967, ApJ, 150, 1005

\bibitem[{Hartle \& Thorne(1968)}]{Hartle:1968si}
Hartle, J.~B., \& Thorne, K.~S. 1968, ApJ, 153, 807

\bibitem[{Haskell {et~al.}(2014)Haskell, Ciolfi, Pannarale, \&
  Rezzolla}]{Haskell:2013vha}
Haskell, B., Ciolfi, R., Pannarale, F., \& Rezzolla, L. 2014, Mon. Not. Roy.
  Astron. Soc., 438, L71

\bibitem[{Hebeler {et~al.}(2013)Hebeler, Lattimer, Pethick, \&
  Schwenk}]{Hebeler:2013nza}
Hebeler, K., Lattimer, J.~M., Pethick, C.~J., \& Schwenk, A. 2013, ApJ, 773, 11

\bibitem[{Hu {et~al.}(2020)Hu, Kramer, Wex, Champion, \& Kehl}]{Hu:2020ubl}
Hu, H., Kramer, M., Wex, N., Champion, D.~J., \& Kehl, M.~S. 2020, Mon. Not.
  Roy. Astron. Soc., 497, 3118

\bibitem[{Kramer {et~al.}(2021)}]{Kramer:2021jcw}
Kramer, M., {et~al.} 2021, Phys. Rev. X, 11, 041050

\bibitem[{Lattimer \& Prakash(2001)}]{Lattimer:2000nx}
Lattimer, J.~M., \& Prakash, M. 2001, ApJ, 550, 426

\bibitem[{Lattimer \& Schutz(2005)}]{Lattimer:2004nj}
Lattimer, J.~M., \& Schutz, B.~F. 2005, Astrophys. J., 629, 979

\bibitem[{Lau {et~al.}(2010)Lau, Leung, \& Lin}]{Lau:2009bu}
Lau, H.~K., Leung, P.~T., \& Lin, L.~M. 2010, ApJ, 714, 1234

\bibitem[{Lau {et~al.}(2017)Lau, Leung, \& Lin}]{Lau:2017qtz}
Lau, S.~Y., Leung, P.~T., \& Lin, L.~M. 2017, Phys. Rev. D, 95, 101302

\bibitem[{Lau {et~al.}(2019)Lau, Leung, \& Lin}]{Lau:2018mae}
---. 2019, Phys. Rev. D, 99, 023018

\bibitem[{Lo \& Lin(2011)}]{Lo:2010bj}
Lo, K.-W., \& Lin, L.-M. 2011, ApJ, 728, 12

\bibitem[{Luk \& Lin(2018)}]{Luk:2018xmt}
Luk, S.-S., \& Lin, L.-M. 2018, ApJ, 861, 141

\bibitem[{Morsink {et~al.}(2007)Morsink, Leahy, Cadeau, \&
  Braga}]{Morsink:2007tv}
Morsink, S.~M., Leahy, D.~A., Cadeau, C., \& Braga, J. 2007, ApJ, 663, 1244

\bibitem[{Mueller \& Serot(1996)}]{Mueller:1996pm}
Mueller, H., \& Serot, B.~D. 1996, Nucl. Phys. A, 606, 508

\bibitem[{Pappas \& Apostolatos(2014)}]{Pappas:2013naa}
Pappas, G., \& Apostolatos, T.~A. 2014, Phys. Rev. Lett., 112, 121101

\bibitem[{Prakash {et~al.}(1988)Prakash, Ainsworth, \&
  Lattimer}]{Prakash:1988md}
Prakash, M., Ainsworth, T.~L., \& Lattimer, J.~M. 1988, Phys. Rev. Lett., 61,
  2518

\bibitem[{Shao \& Yagi(2022)}]{Shao:2022koz}
Shao, L., \& Yagi, K. 2022, Sci. Bull., 67, 1946

\bibitem[{Silva {et~al.}(2021{\natexlab{a}})Silva, Holgado,
  C\'ardenas-Avenda\~no, \& Yunes}]{Silva:2020acr}
Silva, H.~O., Holgado, A.~M., C\'ardenas-Avenda\~no, A., \& Yunes, N.
  2021{\natexlab{a}}, Phys. Rev. Lett., 126, 181101

\bibitem[{Silva {et~al.}(2021{\natexlab{b}})Silva, Pappas, Yunes, \&
  Yagi}]{Silva:2020oww}
Silva, H.~O., Pappas, G., Yunes, N., \& Yagi, K. 2021{\natexlab{b}}, Phys. Rev.
  D, 103, 063038

\bibitem[{Stergioulas \& Friedman(1995)}]{Stergioulas:1994ea}
Stergioulas, N., \& Friedman, J.~L. 1995, ApJ, 444, 306

\bibitem[{Wiringa {et~al.}(1988)Wiringa, Fiks, \& Fabrocini}]{Wiringa:1988tp}
Wiringa, R.~B., Fiks, V., \& Fabrocini, A. 1988, Phys. Rev. C, 38, 1010

\bibitem[{Witten(1984)}]{Witten:1984rs}
Witten, E. 1984, Phys. Rev. D, 30, 272

\bibitem[{Yagi {et~al.}(2014)Yagi, Kyutoku, Pappas, Yunes, \&
  Apostolatos}]{Yagi:2014bxa}
Yagi, K., Kyutoku, K., Pappas, G., Yunes, N., \& Apostolatos, T.~A. 2014, Phys.
  Rev. D, 89, 124013

\bibitem[{Yagi \& Yunes(2013{\natexlab{a}})}]{Yagi:2013awa}
Yagi, K., \& Yunes, N. 2013{\natexlab{a}}, Phys. Rev. D, 88, 023009

\bibitem[{Yagi \& Yunes(2013{\natexlab{b}})}]{Yagi:2013bca}
---. 2013{\natexlab{b}}, Science, 341, 365

\bibitem[{Yagi \& Yunes(2015)}]{Yagi:2015hda}
---. 2015, Phys. Rev. D, 91, 123008

\end{thebibliography}
\bibliographystyle{aasjournal}

\end{document}